\documentclass[12pt]{article}
\usepackage{sc3conf}
\usepackage{amsfonts}
\usepackage{epsfig}
\usepackage{graphicx}

\def\v#1,{{\bf#1},}
\def\j#1{{\it#1\/}}  
\def\mincir{\raise -2.truept\hbox{\rlap{\hbox{$\sim$}}\raise5.truept
\hbox{$<$}\ }}
\def\magcir{\raise -4.truept\hbox{\rlap{\hbox{$\sim$}}\raise5.truept
\hbox{$>$}\ }}

\begin{document}
\raggedbottom

\title{Young compact objects in the Solar vicinity}

\authors{S.B.~Popov,\adref{1,4}
  M.E.~Prokhorov,\adref{1}
  M.~Colpi,\adref{2}
  A.~Treves,\adref{3}  
  and R.~Turolla\adref{4}}

\addresses{\1ad Sternberg Astronomical Institute,
                 Universitetski pr. 13, Moscow 119992, Russia,
  \nextaddress \2ad University of Milano-Bicocca, 
                     Piazza della Scienza 3, Milano 20126, Italy,
  \nextaddress  \3ad University dell'Insubria,
                      via Vallegio 11, Como 22100, Italy,
  \nextaddress \4ad University of Padova,
                    via Marzolo 8, Padova 35131,  Italy}

\maketitle

\begin{abstract}
We present Log N -- Log S distribution for close-by young isolated neutron
stars. On the basis of this distribution it is shown
that the seven ROSAT isolated neutron stars (if they are
young cooling objects) are genetically related to the Gould Belt.
We predict, that there are about few tens unidentified
close-by young isolated neutron stars in the ROSAT All-Sky Survey.
The possibility that these seven peculiar sources contain a neutron 
star less massive and more magnetized then in ordinary radiopulsars is 
also discussed.
In the aftermath of relatively close recent supernova explosions
(1 kpc around the Sun, a few Myrs ago),    
a few black holes might have been formed,    
according to the local initial mass function.
We thus  discuss the possibility of determining  approximate positions
of close-by isolated black holes using  data on runaway stars
and simple calculations of binary evolution and disruption.
\end{abstract}

\section{Introduction}

Neutron stars (NSs) and black holes (BHs)
are among the most interesting astrophysical
sources. Usually NSs are observed as radio pulsars or as accreting
objects in close binaries. Similarly, stellar mass BHs are observed
when they accrete matter from a companion star.
Here we focus on much more elusive sources, namely isolated NSs 
(which may show no radiopulsar activity) and isolated BHs.

An isolated NS can be relatively bright in soft X-rays
due to its thermal emission
during the first Myrs of its life, when it is still hot ($T\sim 10^{6}$~K)
in the aftermath of the supernova (SN) explosion.
Such objects are observed in the Solar proximity
and in SN remnants \cite{bp}. Older NSs (that is to say
those which crossed the deathline in $\approx 10^7$ yr) are not expected to 
emit appreciable amounts of electromagnetic radiation in any energy band.
However, accretion of the interstellar medium (ISM) may make them
shine again as soft, faint X-ray sources (see e.g. Treves et al. 
\cite{pasp}). Much in the same way, 
an isolated BH may be detected if it accretes from the ISM, or, possibly,
revealed through microlensing \cite{agol}.

In this paper we discuss the possible origin of close-by, isolated NSs
and present some evidence that the seven radio-quiet, thermally emitting
ROSAT sources (the ``magnificent seven'' \cite{pasp}) may be characterized
by different values of the stellar parameters (mass and magnetic field) 
with respect to ordinary radiopulsars.
We first construct the Log N -- Log S distribution for young 
close-by isolated NSs and compare it with present observations 
of close-by young isolated NSs of all types.  
Then we discuss how the alignment of the magnetic and rotation axes, together
with the role of fall-back following the supernova event, may help in
explaining the observed parameters of NSs.
Finally in \S 3 we discuss how one can estimate an approximate positions 
of close-by young isolated BHs in the light of their possible detection
at X/gamma-rays energies.

\section{Isolated neutron stars}

In this section we discuss isolated NSs. The material presented here is 
partly based on the results published in Popov et al. \cite{p02}. In addition
some recent results are also included \cite{p03}.

\subsection{Origin of close-by isolated NSs}

To understand how the local population of isolated NSs originated, we
construct their Log N -- Log S distribution. 
The main components of our model are (see \cite{p03}): 
spatial distribution of NS progenitors,
NS formation rate, NS cooling history, 
and a model of interstellar absorption
(that is to say the spatial distribution of the ISM).
In addition we calculate the dynamical evolution of NSs in the galactic
potential. In brief our model can be described in the following way:
NSs are born in the Galactic plane and in the Gould Belt (a local
compound of stellar associations, see below);
at birth they receive a kick velocity; we then follow the evolution
of NSs in the Galactic potential; finally, we calculate the 
ROSAT count rate basing on cooling curves and an assumed model of 
interstellar absorption.

NSs are considered to be born with a constant rate:
20 NSs per Myr come from the Gould Belt,
and 250 NSs per Myr from the Galactic plane
(up to a limiting distance of 3 kpc from the Sun) with a uniform 
distribution. The Gould Belt is modeled as a disk of 500 pc radius with an 
inclination of 18$^\circ$ with respect to the Galactic plane.
Its center is situated at 100 pc from the Sun in the 
Galactic anticenter direction. The central region ($150$ pc in radius)
is devoided of newborn NSs (see P\"oppel \cite{poppel}  
and Torra et al. \cite{torra}).

\begin{figure}[ht]
\begin{center}
    \epsfig{file=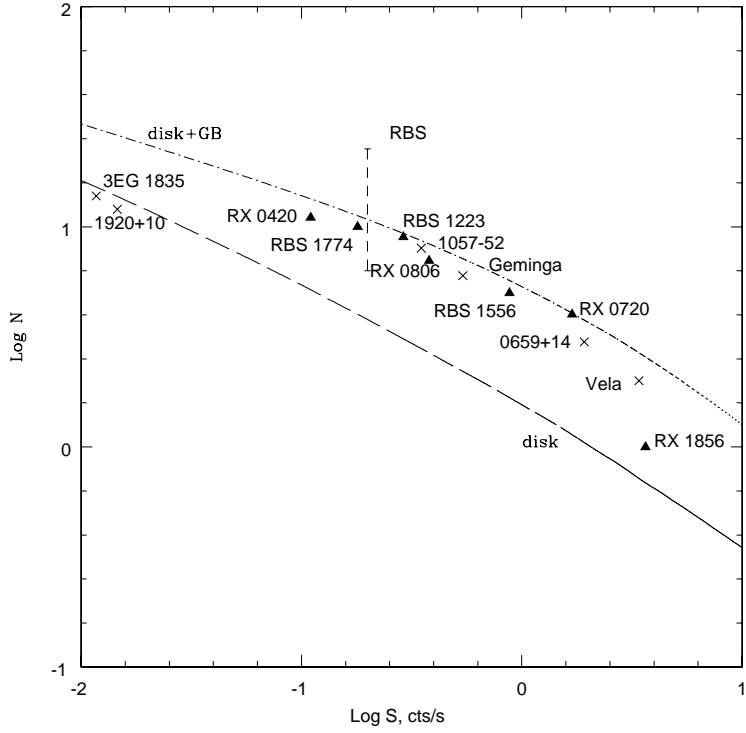, bb = 18 144 592 718, width=0.75\textwidth} 
    \caption{All-sky Log N - Log S distribution.
Black triangles -- the seven RINSs;
crosses -- Geminga, ``three musketeers'', 1929+10 and 3EG J1835.
We also show the ROSAT Bright Sources
(RBS) limit (Schwope et al. \cite{Schwope}).
Upper curve: NSs born in the Gould Belt and in the Galactic disk
($r_{\rm disk}=3$ kpc, total birth rate 270 Myr$^{-1}$).
Lower curve: NSs born only in the Galactic
disk ($r_{\rm disk}=3$ kpc, birth rate 250 Myr$^{-1}$).}\label{fig:1}
  \end{center}
\end{figure}

To calculate the thermal evolution of NSs we use the data obtained by
Sankt-Petersburg group (see Kaminker et al. \cite{kam},
and the review by Yakovlev et al. \cite{yak99}). The NS cooling depends
on the star mass and we adopt a flat mass spectrum in the range
$1.1\, M_{\odot} < M < 1.8\, M_{\odot}$.
A more standard spectrum with a sharp maximum around 1.35-1.4 $M_\odot$
gives nearly the same result. Cooling curves take into account all neutrino 
processes but ignore neutron superfluidity in the crust and core since
this is not expected to influence the final results significantly.
Calculations for each NS are truncated when its temperature drops to
$10^5$ K; this corresponds to a NS age of 4.25 Myrs for the lightest NSs
($M=1.1\, M_{\odot}$) or less for more massive objects.

Since we expect the NS to emit most of its luminosity at UV/soft X-ray
energies ($\sim 20-200$ eV or $T\approx 10^5$--$10^6$ K)
interstellar absorption plays a crucial role as far as the observability
of these sources is concerned.
Any attempt to estimate the amount of observable cooling isolated NSs
using unabsorbed flux {\it greatly} overestimates their number.

Our main results are presented in Fig.~1 where we 
compare the Log N -- Log S for NSs born in the Gould Belt
and the Galactic disk. All curves refer to the whole sky. 
As can be seen the contribution from NSs born in the disk fails to explain 
the observed distribution while the inclusion of the objects originating from
the Gould Belt {\it alone} can match the observations.
Absorption, the flat geometry of NS initial distribution and 
the finite extension of the Belt naturally 
explain the very flat (slope $<-1$)  Log N -- Log S distribution. 

Our calculations show that there may be at most
a few dozens of unidentified
close-by isolated NSs in the ROSAT All-Sky Survey (at count rate
$>$ 0.015 cts s$^{-1}$)
depending on parameters of the model.
Also there may be a few unidentified ROSAT isolated NSs
(RINSs) with fluxes $> 0.1$ cts s$^{-1}$
at low Galactic latitudes
(see also Schwope et al. \cite{Schwope}). Most objects should be observed  
at $|b|< $ 20$^\circ$ towards the directions
of lower absorption.
Some of them can have counterparts among unidentified gamma-ray sources
(also possibly connected with the Gould Belt, see Grenier \cite{Gr}).
Identification of these objects 
can be important for choosing a correct cooling model and 
for determination of the mass spectrum of NSs.

\subsection{Census of close-by young NSs}
 
At present about 20 NSs are known which are 
younger than 4.25 Myrs and closer than 1 kpc to the Sun
(see the Table). They include: 
the ``magnificent seven'' (radio-quiet, ROSAT isolated NSs with only
thermal emission), Geminga and
the Geminga-like object 3EG J1835 (pulsars the beams of which 
do not intersect the Earth), the ``three musketeers'' (Vela, PSR 
0656+14, PSR 1055-52), 
PSR 1929+10 and seven 
young radio pulsars, which have not been detected in X-rays yet.

In addition to the observed sources, we 
expect about one hundred isolated NSs younger than $\sim $4 Myr inside 1 kpc.
These NSs are not detected as radio pulsars,
but tens of them can be identified
in ROSAT data as dim sources (others are too old to be hot enough). 
Pulsar beaming can be responsible only for
a fraction  of these undetected (in the radio) young NSs (about 50-70\%
of young pulsars are not visible from Earth \cite{tm}),
and most of RINSs should be really radio silent.
This provides strong support to the arguments by Gotthelf and Vasisht 
\cite{gv}, that ``at least half of the observed young neutron stars follow an
evolutionary path quite distinct from that of the Crab pulsar''.

\begin{table}
\caption{Local ($r<1$ kpc) population of young (age $<4.25$ Myrs)
isolated neutron stars.\vspace*{1pt}}
{\footnotesize
\begin{tabular}{|l||c|c|c|c|c|c|}
\hline
\hline

{} & {} & {} &{} &{} &{} &{}\\[-1.5ex]
Object name & Period, & CR$^a$, & $\dot P$ &Dist.,& Age$^b$, & Ref.
\\[1ex]
            &   s     & cts/s & $/10^{-15}$& kpc   & Myrs     &
\\[1ex]
\hline
 & & & & & & \\[1ex]
RX J1856.5-3754            &  ---  & 3.64  &  ---&0.117$^e$&$\sim0.5$&
\cite{pasp,kaplan}\\[1ex]
RX J0720.4-3125                 & 8.37 & 1.69  &$\sim 30-60$& ---&---&
\cite{pasp,zane}\\[1ex]
RX J1308.6+2127 & 10.3 & 0.29  & $< 10^{4} ?$&---&---&
\cite{pasp,hamb,haberl}\\[1ex]
RX J1605.3+3249       &  ---  & 0.88  & --- & --- & --- &
\cite{pasp}\\[1ex]
RX J0806.4-4123                 &  11.37  & 0.38  & --- &---&--- &
\cite{pasp,hz}\\[1ex]
RX J0420.0-5022                 &  22.7   & 0.11  & --- &---   &--- &
\cite{pasp}\\[1ex]
RX J2143.7+0654 &  ---    & 0.18  & --- & ---  &--- &
\cite{zamp}\\[1ex]
\hline
 & & & & & & \\[1ex]
PSR B0633+17            & 0.237 & 0.54$^d$ &10.97&0.16$^e$&0.34&
\cite{bt}\\[1ex]
3EG J1835+5918 & ---   & 0.015    & --- & ---  &  --- &
\cite{mira}\\[1ex]
\hline
 & & & & & & \\[1ex]
PSR B0833-45    & 0.089 & 3.4$^d$  & 124.88 & 0.294$^e$ &
0.01&
\cite{bt,pavlov,atnf}\\[1ex]
PSR B0656+14          & 0.385 & 1.92$^d$ &  55.01 & 0.762$^f$ & 0.11
&
\cite{bt,atnf}\\[1ex]
PSR B1055-52          & 0.197 & 0.35$^d$ &   5.83 & $\sim 1^c$ &
0.54&
\cite{bt,atnf}\\[1ex]
PSR B1929+10          & 0.227 & 0.012$^d$& 1.16 &  0.33$^e$  & 3.1&
\cite{bt,atnf}\\[1ex]
\hline
 & & & & & & \\[1ex]
PSR J0056+4756  & 0.472 & --- & 3.57 &  0.998$^f$  & 2.1& 
\cite{atnf}\\[1ex]
PSR J0454+5543  & 0.341 & --- & 2.37 &  0.793$^f$  & 2.3& 
\cite{atnf}\\[1ex]
PSR J1918+1541  & 0.371 & --- & 2.54 &  0.684$^f$  & 2.3& 
\cite{atnf}\\[1ex]
PSR J2048-1616  & 1.962 & --- & 10.96&  0.639$^f$  & 2.8& 
\cite{atnf}\\[1ex]
PSR J1848-1952  & 4.308 & --- & 23.31&  0.956$^f$  & 2.9& 
\cite{atnf}\\[1ex]
PSR J0837+0610  & 1.274 & --- & 6.8  &  0.722$^f$  & 3.0& 
\cite{atnf}\\[1ex]
PSR J1908+0734  & 0.212 & --- & 0.82 &  0.584$^f$  & 4.1& 
\cite{atnf}\\[1ex]
\hline
\hline
\multicolumn{7}{l}{
$^a$ ROSAT count rate}\\
\multicolumn{7}{l}{
$^b$) Ages for pulsars are estimated as $P/(2\dot P)$,}\\
\multicolumn{7}{l}{
      for RX J1856 the estimate of its age comes from kinematical
      considerations.}\\
\multicolumn{7}{l}{  
$^c$) Distance to PSR B1055-52 is uncertain ($\sim$ 0.9-1.5 kpc)}\\
\multicolumn{7}{l}{  
$^d$) Total count rate (black body + non-thermal)}\\
\multicolumn{7}{l}{  
$^e$) Distances determined through parallactic measurements}\\
\multicolumn{7}{l}{  
$^f$) Distances determined with dispersion measure}\\
\hline
\end{tabular} }
\vspace*{-13pt}
\end{table}

\subsection{Are RINSs of a different stock ?}

An interesting feature of RINSs population is the detection of periods
in the $\sim$~10-20 s range (typical of SGRs/AXPs, i.e. of magnetars)
for four objects. Present data allow to exclude any pulsation 
(down to a few \% fraction) for at least one source, RX J1856.5-3754.
V. Beskin (2001, private communication) suggested this could be due 
to the alignment of magnetic and spin axes (see for example
Tauris \& Manchester \cite{tm} for a discussion).
Alignment is a process which leads to  ``period freezing'' and
a low pulsed fraction.

However in the case of coolers
alignment should operate on short timescale, since the star cools down
in  $\approx 1$ Myr. For radio pulsars the timescale of alignment is 
about 10 Myrs or longer \cite{tm}, so it seems unlikely that
this mechanism is responsible for RINSs distribution of the pulsed
fraction, unless one assumes that RINSs form a separate population from
normal radio pulsars. To illustrate this let us assume that the alignment 
timescale is 
$\tau_{align} \propto (\Omega_0^2 {\rm cos}^2 \alpha_0 \, B_0^2)^{-1}$
(here $\alpha_0$ and $B_0$ are initial values of an angle between spin and
magnetic axis and of magnetic field).
The previous expression comes from magnetodipolar braking supplemented
by the condition $\Omega_0 \, {\rm cos} \alpha_0=\Omega \, {\rm cos} \alpha$.
To explain the difference between $\tau_{align}$ in RINSs and radio pulsars
RINSs need to have a different distribution in $B_0$ 
and/or $\alpha_0$. In this respect RINSs may come from the same population
as radiopulsars but are characterized by different average
properties, like e.g. higher values of the magnetic field and
relatively higher surface temperatures. The latter would 
imply a lower mass for NSs of the same age.

RINSs are currently thought to be rather highly magnetized objects
(in RX J0720.4-3125 the detected spind-down implies $B\sim 2\times 10^{13}$ G
\cite{zane}).
If this is indeed the case, then one has to explain why their $B$-field
is a factor $\sim 10$ higher than the average value in radiopulsars
($\sim 2\times 10^{12}$ G). A possibility is that NSs with higher 
magnetic fields are hotter. It is known that less massive NSs cools
more slowly because direct URCA processes are not effective (see e.g.
\cite{yak99}). This means that among NSs of the same age the lighter
are the hotter, so to test our hypotesis we need to show that lighter
NSs may support a stronger field.  
Such a correlation arises quite naturally  
if more massive NSs get their additional mass from fall-back.
In this case their magnetic field can be significantly 
suppressed \cite{page}, so more massive NSs should have lower initial 
magnetic fields. 
Besides, strong initial magnetic field together with fast rotation
can prevent strong fall-back
(this is especially possible if the magneto-rotational
mechanism of supernova explosion is valid,
see \cite{bk,pp01}). Again this leads to the same correlation
between mass and filed strength, i.e. NSs with stronger fields 
would have lower masses. Also, this picture makes room for long initial 
spin periods.
The study of this (and possible other) correlation
in isolated NSs may prove very useful in understanding the correct
mechanism of the SN event which gave birth to these objects. 

In summarizing \S 2, we stress that future determination of RINSs 
parallax and proper motion may help in
tracing back their kinematical history and derive their age.
It can give a clue to their mass determination basing on cooling curves  
(see Kaminker et al. \cite{kam}). Our results suggest that the fraction of
low-mass NSs ($M\mincir$1.3 $M_{\odot}$) may not be small. 
On the other hand there should be room for  NSs with $M\magcir 1.4 M_\odot$, 
because otherwise the number of bright objects would be too large.

\section{Close-by young isolated black holes}

In this section we base on the results published in Popov et
al. \cite{p02}
and Prokhorov and Popov \cite{pp02}.

SNae explosions produce not only NSs, but also BHs.
It is commonly accepted that BHs are one order of magnitude less abundant
than NSs. This estimate comes from the critical mass for BH formation and
follows if one assumes that progenitors more massive than about 
35~$M_{\odot}$ ended as BHs.
Having dozens of SNae in the close solar vicinity during the last 10 Myr
we can expect several BHs to have formed during the same period in the
solar neighborhood.

At present, 56 runaway stars are known within $\sim$~700 pc from 
the Sun\cite{10}. Only a few of them result from star-star interactions, so
the majority comes from SNae explosions in binary systems.
If the above considerations are correct we can expect about 5 BHs
formed in about 50 disrupted binaries.

Close-by massive runaway stars give us a chance to calculate the
approximate position of close-by young isolated BHs.   
Among runaway stars the most massive are
$\lambda$ Cep, $\zeta$ Pup, HIP 38518 and  $\xi$ Per (see 
Hoogerwerf et al. \cite{10}). Since their mass is  
$\magcir 33$ $M_{\odot}$, the
companion (actually the {\it primary} in the original binary) was even more
massive on the main sequence stage. So, the most likely product
of the explosion of such a massive star  should be a BH.

If the present velocities
of runaway stars are known, one can estimate
their ages and places of birth. This has been done by Hoogerwerf et
al. \cite{10}.
To calculate the present position of a BH we have to know the binary  
parameters, i.e., the 
masses of stars before the explosion, the BH mass, the  eccentricity
of the orbit before the explosion, the
orbit orientation, and finally the kick velocity of the BH. 
Some parameters can be inferred from the  observation of the
secondary star. We can assume a zero kick velocity  for  BHs and
zero orbital eccentricity. 
Other parameters should be varied within assumed ranges
(see details in Prokhorov and Popov \cite{pp02}.

We calculated approximate positions of isolated BHs for the four systems
mentioned above and estimated error boxes where these BHs could be found. 
For $\xi$ Per and $\zeta$ Pup we obtained not very large
error boxes inside each of which only one unidentified EGRET source is known.
We suggest that these  objects can be young isolated BHs.
For the two other systems (HIP 38518 and $\lambda$ Cep) the present 
position of the BH is more uncertain and no definite conclusion on the
possible detectability of the collapsed object can be drawn.

\section{Conclusions}

We have presented evidence that the seven radio-quiet ROSAT isolated 
NSs discovered so far can be connected with
recent SNae explosions in the Gould Belt. These events produced 
nearby runaway stars and peculiar features in the local ISM 
including the Local Bubble. The relatively high
local spatial density of young NSs is a natural consequence of the 
large number of massive progenitors in the Belt. The lack of a similar
overabundance of active radiopulsars in the Solar vicinity lends further
support to the claim that a large fraction ($\sim 50\%$ or more)   
of young NSs should be radio quiet.   
According to our results, the 
ROSAT All Sky Survey may contain about a few tens of unidentified RINSs.
Moreover, it is possible that some unidentified RINSs with quite large
flux ($> 0.1$ cts s$^{-1}$) are still hiding at low Galactic latitudes.

We also propose that massive runaway stars may be used to trace
the present position of young close-by isolated BHs. Our calculations allowed
to estimate with reasonable accuracy the positions of four such BHs. 
In two cases the error box is not too large and this may lead in the future
to the positive identification of an isolated BH with X/gamma-rays 
observations.

\section{Acknowledgments}

We want to thank D.G. Yakovlev for the data on cooling curves
and comments on them, and
V.S. Beskin, M. Chieregato, A. Possenti 
and L. Zampieri for discussions.
The work of S.P. was supported by the RFBR grant 02-02-06663
and by RSCI; that of M.P.by RFBR grant 01-15-99310.
S.P. thanks Universities dell'Insubria and Milano-Bicocca for hospitality.

\end{document}